# Universal superconducting fluctuations and the implications for the phase diagram of the cuprates

G. Yu[1], D.-D. Xia[1,2], N. Barišić[1,3], R.-H. He[4], N. Kaneko[5], T. Sasagawa[6], Y. Li[7], X. Zhao[1,2], A. Shekhter[8], and M. Greven[1]

[1] School of Physics and Astronomy, University of Minnesota, Minneapolis, Minnesota 55455, USA
[2] State Key Lab of Inorganic Synthesis and Preparative Chemistry, College of Chemistry, Jilin University, Changchun 130012, China
[3] Service de Physique de l'Etat Condensé, CEA-DSM-IRAMIS, F 91198 Gif-sur-Yvette, France
[4] Advanced Light Source, Lawrence Berkeley National Laboratory, Berkeley, California 94720, USA
[5] National Institute of Advanced Industrial Science and Technology (AIST), 1-1-1 Umezono, Tsukuba, Ibaraki 305-8563, Japan
[6] MSL, Tokyo Institute of Technology, Kanagawa 226-8503, JAPAN
[7] International Center for Quantum Materials, School of Physics, Peking University, Beijing 100871, China
[8] Pulsed Field Facility, NHMFL, Los Alamos National Laboratory, Los Alamos, NM 87545

**Superconductivity in the cuprates emerges from an enigmatic metallic state. There remain profound open questions regarding the universality of observed phenomena and the character of precursor fluctuations above the superconducting (SC) transition temperature ($T_c$). For single-CuO$_2$-layer La$_{2-x}$Sr$_x$CuO$_4$ (LSCO) and Bi$_2$(Sr,La)$_2$CuO$_{6+\delta}$ (Bi2201), some experiments[1-5] seem to indicate an onset of SC fluctuations at very high temperatures (2-3 times $T_c^{max}$, the $T_c$ value at optimal hole concentration $p$), whereas other measurements suggest that fluctuations are confined to the immediate vicinity of $T_c(p)$[6-9]. Here we use torque magnetization to resolve this conundrum by systematically studying LSCO, Bi2201 and HgBa$_2$CuO$_{4+\delta}$ (Hg1201). The latter is a more ideal single-layer compound[10], featuring high structural symmetry, minimal disorder, and $T_c^{max}$ = 97 K, a value more than twice those of LSCO and Bi2201. We find in all three cases that SC diamagnetism vanishes in an unusual exponential fashion above $T_c$, and at a rapid rate that is universal. Furthermore, the high characteristic fluctuation temperatures of LSCO and Bi2201 closely track $T_c(p)$ of Hg1201. These observations suggest that, rather than being indicative of SC diamagnetism, the fluctuations at high temperatures in the low-$T_c^{max}$ compounds are associated with a competing order. This picture is further supported by an analysis of available results for double-layer cuprates.**

The order parameter of a superconductor is comprised of an amplitude and a phase, and the SC state requires phase coherence over macroscopic distances. In conventional metals, phase stiffness (i.e., the propensity to lock the phase over macroscopic distances) is large, and superconductivity sets in as soon as the pair amplitude becomes non-zero[11]. It has been argued that the phase stiffness in the cuprates is softened by non-local electron-electron interactions[12], and that the pair amplitude remains nonzero in the metallic 'normal' state well above $T_c$, even though global phase coherence cannot be maintained[12,13]. In this unconventional 'preformed-pair' scenario, the metallic state above $T_c$ is expected to exhibit strong fluctuations over an extended temperature range. Supporting these ideas, numerous experiments (Nernst effect[1,2], torque magnetization[3,4], photoemission[5] and magnetoresistance[14]) have been interpreted as indicative of an extended fluctuation regime. However, in other experiments (microwave[6], teraherz conductivity[7,8] and specific heat[9]), signatures of SC fluctuations have been discerned only in a relatively narrow temperature range above $T_c$, which invites a more conventional interpretation. In order to advance the field, it is essential to separate universal from nonuniversal phenomena in the unusual metallic state.

At moderate and intermediate hole concentrations ($p$), the cuprates enter the 'pseudogap' phase below the temperature $T^*$, in which an energy gap develops on a considerable portion of the two-dimensional Fermi surface. For underdoped samples, $T^*(p)$ is much larger than $T_c(p)$. The primary focus of the present work is on the model compound Hg1201, which features a simple tetragonal crystal structure and the highest value of $T_c^{max}$ among all single-$CuO_2$-layer compounds[10,15,16]. For Hg1201 and several double-layer compounds, it has been established that the pseudogap region of the phase diagram is a genuine new phase of matter, characterized by a distinct magnetism[17,18], consistent with prior theoretical work[19]. It is intriguing that the reports of an extended SC fluctuation regime pertain to low-$T_c^{max}$ systems[1,2,5] such as LSCO, where disorder effects are known to be more prominent[10] and the pseudogap magnetism is not fully developed[20].

Among the experimental techniques that have been applied to the study of SC fluctuations in the metallic state of the cuprates, magnetometry is unique, for it directly probes diamagnetism, a highly distinctive characteristic of superconductivity[21]. Moreover, unlike complex transport and relaxation phenomena, magnetization is a thermodynamic quantity and hence directly related to the equilibrium (thermodynamic) state of a system. In torque magnetometry, the magnetization is deduced from the mechanical torque experienced by a sample in an external magnetic field. A torque, $\tau = V\mu_0(\mathbf{M} \times \mathbf{H})$, is created when the sample magnetization (**M**) is not collinear with the applied magnetic field (**H**). Here, $\mu_0$ is the permeability of free space and $V$ is the sample volume. The measured torque is a function of temperature ($T$), magnetic field strength ($H$), and orientation of the sample with respect to the direction of the field. The sample orientation is parameterized by the angle $\theta$ between **H** and the crystallographic $c$-axis. We analyze the torque susceptibility $\chi_{torque}(H,\theta,T)$ and torque magnetization $M_{torque}(H,\theta,T)$ defined via $\chi_{torque} \equiv M_{torque}/H \equiv \tau/(V\mu_0 H_a H_c)$, where $H_a = H\sin(\theta)$ and $H_c = H\cos(\theta)$ are the components of **H** along the crystallographic $a$ and $c$ directions. At high temperature, the magnetization is dominated by a linear-in-field paramagnetic response, $\mathbf{M}_p(\mathbf{H},T) = \chi_p(T)\mathbf{H}$, and $\chi_{torque}$ equates to the linear susceptibility difference $\chi_c(T) - \chi_a(T)$, which is independent of field strength and sample orientation. Near $T_c$, $M_{torque}$ is a non-linear function of $H$, and $\chi_{torque} = M_c/H_c - M_a/H_a$ is a useful measure of SC diamagnetism, since the magnetization (and the torque) in the highly anisotropic cuprates is dominated by the out-of-plane component $M_c$.

In prior torque studies of LSCO and Bi2201[3,4], it was suggested that diamagnetism due to SC fluctuations can be inferred from the deviation $\mathbf{M}_d = \mathbf{M} - \mathbf{M}_p$, where the paramagnetic background $\mathbf{M}_p(\mathbf{H},T) = (a + bT)\mathbf{H}$ was estimated from the high-temperature behavior. As shown for Hg1201 in Fig. 1(a) (and Figs. S2&S3 for Bi2201 and LSCO), the assumption of a $T$-linear magnetization over a wide temperature range is unreliable. We note that at temperatures well above $T_c$, $\chi_a$ and $\chi_c$ may be affected by phenomena such as the opening of the pseudogap[17-19] and the emergence of other electronic phases[22]. A more reliable signature of SC diamagnetism in the metallic state is derived from its magnetic field dependence, which is expected to be non-linear in field even at relatively low fields (see Supplementary Information (SI)). Torque magnetization allows the determination of the field dependence of magnetization either directly, through field scans at fixed angle, or indirectly, by observing its angular dependence at fixed field strength. In the angular scans, the linear-in-field magnetization (paramagnetic or diamagnetic) reveals itself as the second harmonic in the angular dependence, $\tau \propto H_a H_c \propto \sin(2\theta)$, whereas the non-linear-in-field magnetization introduces higher harmonics. For example, for our $T_c \approx 96$ K Hg1201 sample, upon approaching $T_c$, the non-linear magnetization in the field scans and the higher harmonics in the angular scans are clearly discernible below the same temperature of about 120 K (Figs. 1b and 1c). These two methods of obtaining $\chi_{\text{torque}}(H_c)$ agree remarkably well (Fig. 1d). The non-linear-in-field component of the magnetization can be quantified by taking the difference ($\Delta_H \chi_{\text{torque}}$) of $\chi_{\text{torque}}$ at two different fields (or, equivalently, at two different angles), since the subtraction removes all linear-in-field components.

At temperatures just above $T_c$, the torque susceptibility of Hg1201 exhibits Ginsburg-Landau-like power-law field dependence at low fields[11] (Fig. 2a & SI). As shown in Figs. 2b&S1, we find that outside of this regime (for fields above ~ 1 T), the diamagnetic signal follows an unusual rapid exponential temperature dependence: $\chi_{\text{torque}} \propto \exp\{-(T-T_c)/T_{\text{fl}}\}$, where $T_{\text{fl}}$ is a measure of the temperature range over which SC fluctuations are significant (Fig. 2c). As demonstrated in Fig. 2d, the exponential dependence is even clearer (over more than two decades in magnitude) from $\Delta_H \chi_{\text{torque}}$, i.e., when the paramagnetic response (dominant at higher temperature) is removed. This unusual behavior is qualitatively consistent with prior magnetization[23] and conductivity[24] measurements.

The data for Hg1201 (Figs. 1&2), Bi2201 (Fig. S2) and LSCO (Fig. S3) reveal that for fields larger than 1 T the magnetization above $T_c$ exhibits two distinct behaviors. The near-$T_c$ regime features the exponential decay of $\chi_{\text{torque}}$ and a strong non-linear field dependence of the magnetization. At higher temperature, $\chi_{\text{torque}}$ exhibits qualitatively different temperature dependence and no (within our sensitivity limit) magnetic field dependence. Consequently, only the near-$T_c$ regime can be confidently ascribed to SC fluctuations, since it directly connects to the SC diamagnetism below $T_c$. The magnetization at higher temperature must have a different physical origin associated with the anomalous normal state.

Figures 3a&b for Hg1201 demonstrate that $T_{\text{fl}}$ has a weak doping dependence down to $p \approx 0.07$, the lowest hole concentration of our study. Remarkably, our complementary measurements of Bi2201 and LSCO reveal that the diamagnetic response of all three compounds is nearly indistinguishable (Figs. 3c, S2, S3), despite the stark difference (a factor of about 2.5) in superconducting transition temperatures. This is summarized in Fig. 3d, which in addition to $T_{\text{fl}}$, obtained directly from $\chi_{\text{torque}}$, also shows a contour

of the nonlinear magnetic response $\log_{10}(\Delta_H \chi_{torque}/\chi^{250}_{torque})$ for Hg1201. We conclude that the SC diamagnetism in the single-layer cuprates is restricted to a universal and rather narrow temperature range that closely tracks the superconducting dome $T_c(p)$. Consequently, we expect any fluctuation phenomena far above $T_c$ in the low-$T_c^{max}$ compounds to be associated with the physics of the unconventional normal state (for example, Nernst effect[1,2] and photoemission experiments[5] for Bi2201 and LSCO suggest anomalies at temperatures that are dramatically higher than $T_c$; the prior torque magnetization results[3,4] are further discussed in the SI; see also Fig. S4). We emphasize that the interpretation of complex transport probes is uncertain without a clear understanding of the underlying metallic state, which is lacking at this time. For example, the Nernst signal can be strongly affected by charge/spin modulations that lead to the reconstruction of Fermi surface[25], and even in more conventional systems is sensitive to the interplay between subtle details of the band structure and relaxation phenomena[26].

As demonstrated in Fig. 4, these high characteristic temperatures for Bi2201 and LSCO are close to $T_c(p)$ of the high-$T_c^{max}$ system Hg1201. Moreover, they appear to be bound by $T_{c,1}$ = 118 K, the highest transition temperature recorded for any single-layer cuprate (Hg1201 at high pressure[27]). The low-$T_c^{max}$ compounds tend to exhibit lower structural symmetry and are known to be considerably more disordered than their high-$T_c^{max}$ counterparts[10] (see SI). Strong disorder coupled with singular electron-electron interactions might enhance charge-ordering tendencies that compete with the superconductivity. This picture is supported by our observations for the double-layer cuprates. Here, the highest recorded SC transition temperature is $T_{c,2}$ = 154 K, observed for $HgBa_2CaCu_2O_{6+\delta}$ at high pressure[27]. Figure S5 shows various results for $Bi_2Sr_2CaCu_2O_{8+\delta}$ ($T_c^{max}$ = 96 K) and $YBa_2Cu_3O_{6+\delta}$ ($T_c^{max}$ = 93 K), which are intermediate-$T_c^{max}$ compounds ($T_c^{max}/T_{c,2}$ ~ 60%, compared to $T_c^{max}/T_{c,1}$ ~ 30% for Bi2201 and LSCO). Indeed, recent X-ray experiments have demonstrated the existence of charge-density-wave correlations below about 150 K in underdoped $YBa_2Cu_3O_{6+\delta}$.[28,29]

Regardless of the origin of this non-universal behavior, we conclude that whereas in the low-$T_c^{max}$ compounds the interpretation of some experimental results is less certain, SC fluctuations in all single-layer cuprates are present in a universally narrow temperature range above $T_c$ and the underlying physics is revealed most clearly in Hg1201. These results place a strong constraint on theoretical models of the unconventional superconductivity in the cuprates.

**Acknowledgements**
We thank N.P. Armitage, K. Behnia, S.A. Kivelson, N.P. Ong and C.M. Varma for valuable comments on the manuscript, and P.A. Crowell for the use of the 9 T PPMS. The 14 T data were obtained with a 14 T PPMS at the Geballe Laboratory for Advanced Materials at Stanford University. The work on Hg1201 was supported by DOE-BES. The work on Bi2201 and LSCO was supported by the NSF and the NSF MRSEC program.

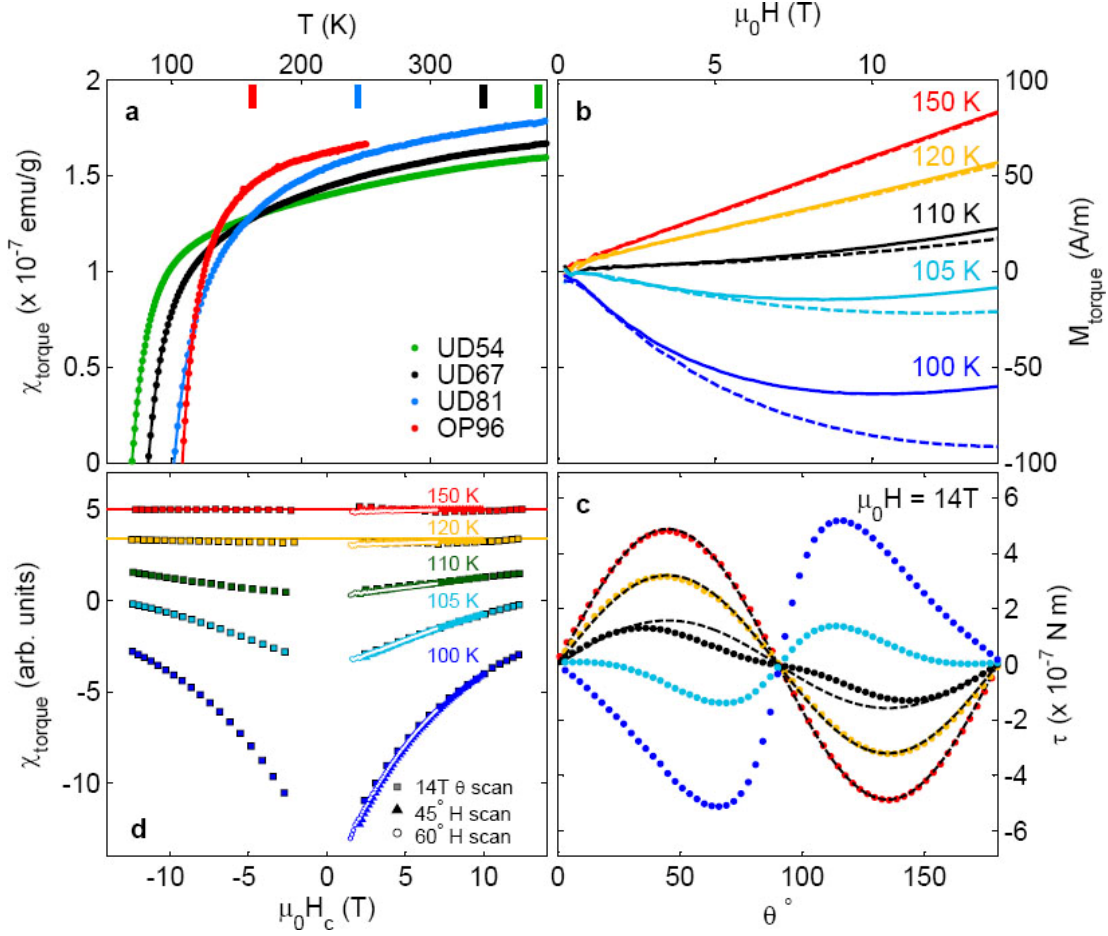

**Figure 1 | Temperature and doping dependence of the susceptibility $\chi_{torque}$ for Hg1201. a**, Torque susceptibility $\chi_{torque}$ (defined in the main text) over a wide temperature range above $T_c$ for three underdoped and one (nearly) optimally doped Hg1201 sample ($T_c \approx$ 54, 67, 81, and 96 K, respectively). $\chi_{torque}$ is obtained with an external magnetic field $\mu_0 H$ = 14 T at $\theta$ = 45°. At fixed field, $\chi_{torque}$ is proportional to the effective magnetization $M_{eff} \equiv \tau/(\mu_0 VH\sin(\theta))$ used in prior torque studies[3,4] (note the difference with our definition of $M_{torque}$). At high temperature, $M_{torque}$ is independent of $\theta$, and $\chi_{torque}$ (now equal to $\chi_c - \chi_a$) is independent of $\theta$ and $H$ up to at least 14 T with very good accuracy. The vertical bars indicate the pseudogap temperature $T^*$ obtained from neutron scattering and planar dc resistivity measurements[6,16-18]. Non-trivial temperature dependences of $\chi_{torque}$ (and $d\chi_{torque}/dT$) are observed up to 400 K. Similar behavior is observed in our complementary measurement of an optimally-doped Bi2201 sample (see SI). **b**, Field dependence of the effective magnetic moment $M_{torque}$ at $\theta$ = 45° (solid lines) and 60° (dashed lines) for sample OP96. Data were taken in 0.2 T intervals. Diamagnetic magnetization is observed below about 120 K, where $M_{torque}(H)$ is strongly nonlinear. **c**, Angular dependence of the torque for sample OP96. The temperature is indicated by the same colors as in **b**. The deviation from the sin(2$\theta$) dependence occurs below the same temperature as the onset of the nonlinear component of $M_{torque}(H)$ in **b**. Dashed black lines are fits to sin(2$\theta$) for 150, 120 and 110 K. **d**, $\chi_{torque}$ as a function of $H_c$ = $H\cos(\theta)$ for OP96, calculated from the field dependence in **b** (triangles and circles) and from the angular dependence in **c** (squares). The two methods agree remarkably well,

indicating that the result is hardly affected by $H_a$. This implies that the contribution from the in-plane response is negligible and that $\chi_c$ dominates the nonlinear diamagnetic signal in this temperature range. Horizontal lines at 120 K and 150 K indicate the field-independent paramagnetic contributions.

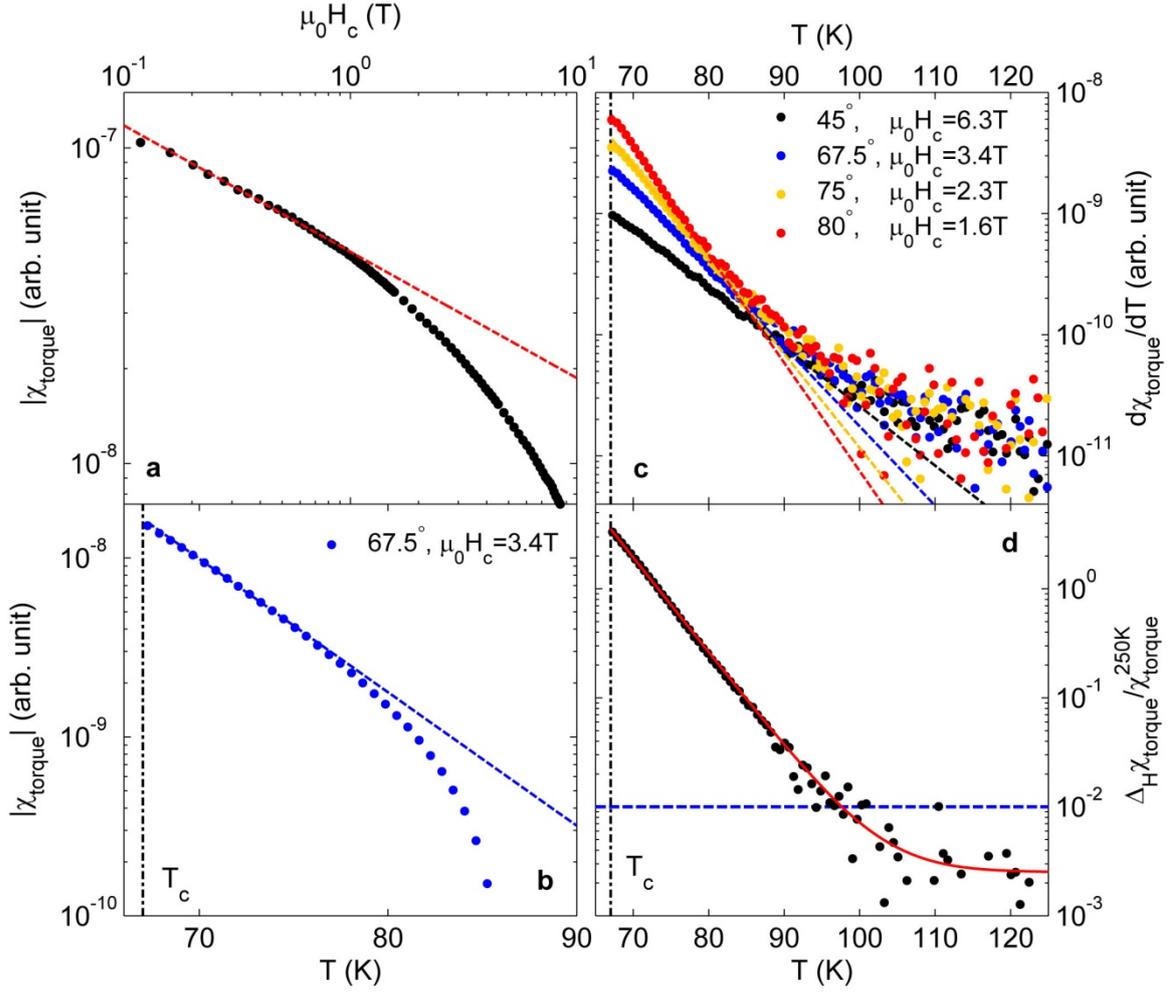

**Figure 2 | Low-field regime, definition of characteristic fluctuation temperature, crossover to high-temperature regime for Hg1201 ($T_c$ = 67 K). a,** Demonstration of power-law (Ginzburg-Laundau-like) field dependence below about 1 T for sample UD67 at 70 K, just above $T_c$. Dashed line represents a fit to $|\chi_{torque}| = |H_c|^{-0.4}$. **b,** Demonstration of exponential temperature dependence $|\chi_{torque}| \propto \exp\{-(T-T_c)/T_{fl}\}$ at a somewhat larger field. Blue dashed line is a fit to the low-temperature data. Vertical dashed-dotted line indicates $T_c$. **c,** The exponential decay is also seen from the derivative $d\chi_{torque}/dT$, which is shown together with fits (lines) at several angles (i.e., different $H_c$). $T_{fl}$ exhibits weak field dependence. Above about $T_c + 15$ K, a crossover to a different behavior is seen. **d,** Difference $\Delta_H\chi_{torque}(T) \equiv \chi_{torque}(45°,T) - \chi_{torque}(67.5°,T)$, normalized by the value at 250 K. The solid red line is a fit to an exponential behavior (with $T_{fl} \approx 5$ K) plus a constant. The blue horizontal dashed line indicates the limit $\Delta\chi_{torque}/\chi^{250}_{torque} = 0.01$ (see SI), below which the diamagnetic signal due to SC fluctuations can no longer be reliably discerned. The data in **b-d** were obtained with a 9 T field.

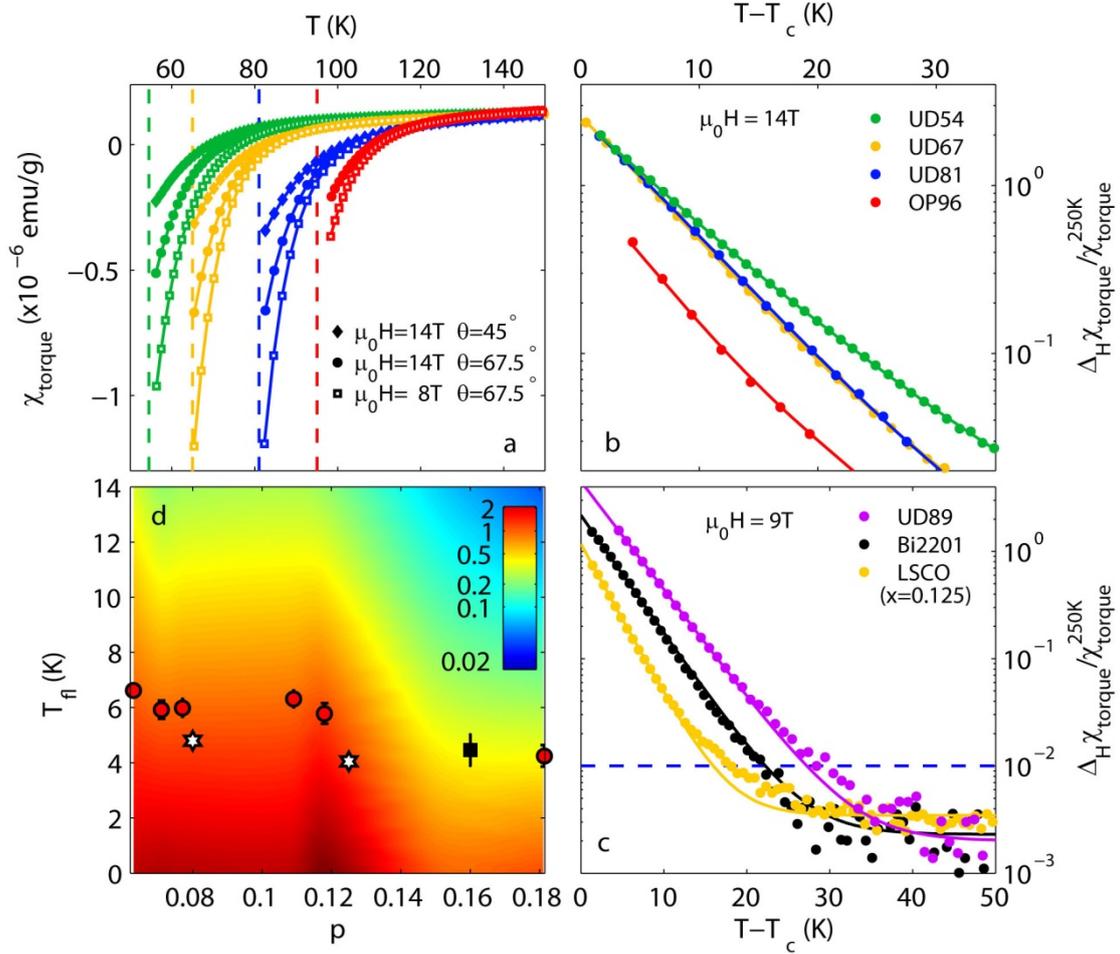

**Figure 3 | Universal behavior of the torque susceptibility. a,** $\chi_{\text{torque}}(H,\theta,T)$ obtained under three different conditions for Hg1201 samples with $T_c$ = 54, 67, 81 and 96 K: (i) $\mu_0 H$ = 14 T, $\theta$ = 45°; (ii) $\mu_0 H$ = 14 T, $\theta$ = 67.5°; (iii) $\mu_0 H$ = 8 T, $\theta$ = 45°. Deviations upon approaching $T_c$ are due to SC fluctuations. Vertical dashed lines indicate $T_c$. $\Delta_H \chi_{\text{torque}}$ can be evaluated as either the difference between (i) and (ii) or between (i) and (iii). **b,** $\Delta_H \chi_{\text{torque}}(T) \equiv \chi_{\text{torque}}(14\text{T},45°,T) - \chi_{\text{torque}}(14\text{T},67.5°,T)$ for Hg1201 (from **a**), normalized at 250 K, versus $T - T_c$. Lines in **a&b** are guides to the eye. **c,** Comparison of $\Delta_H \chi_{\text{torque}} \equiv \chi_{\text{torque}}(9\text{T},67.5°,T) - \chi_{\text{torque}}(9\text{T},67.5°,T)$ for slightly underdoped Hg1201 ($T_c$ = 89 K), optimally doped Bi2201 and LSCO ($x$ = 0.125) (see SI). Solid lines are fits to an exponential behavior plus a constant. Dashed line indicates limit beyond which SC fluctuations can no longer be reliably discerned. **d,** Characteristic SC fluctuation temperature $T_{\text{fl}}$ vs. hole concentration for Hg1201 (circles), Bi2201 (square) and LSCO (stars) extracted for $\mu_0 H_c$ = 3.1-3.4 T from the exponential behavior $\chi_{\text{torque}} \propto \exp\{-(T-T_c)/T_{\text{fl}}\}$ near $T_c$. The errors represent the uncertainty of the fits (see SI). The color contour shows the magnitude of the nonlinear magnetic response $\log_{10}(\Delta_H \chi_{\text{torque}}/\chi^{250}_{\text{torque}})$ for Hg1201 as function of $T-T_c$ and $p$. Normalization of $\Delta_H \chi_{\text{torque}}$ by its high-temperature value allows the comparison of samples with different hole concentrations (see SI). $T_{\text{fl}}$ weakly increases with increasing $H_c$, but this does not affect the observation of universal behavior and the fact that $T_{\text{fl}}$ slightly decreases with increasing hole concentration. While $T_{\text{fl}}$ is best defined from $\chi_{\text{torque}}$, the difference $\Delta_H \chi_{\text{torque}}$ used in Figs. 2-4 and Figs. S2-S4 demonstrates the exponential decay to higher temperature and gives a better estimate of the measurable extent of SC fluctuations.

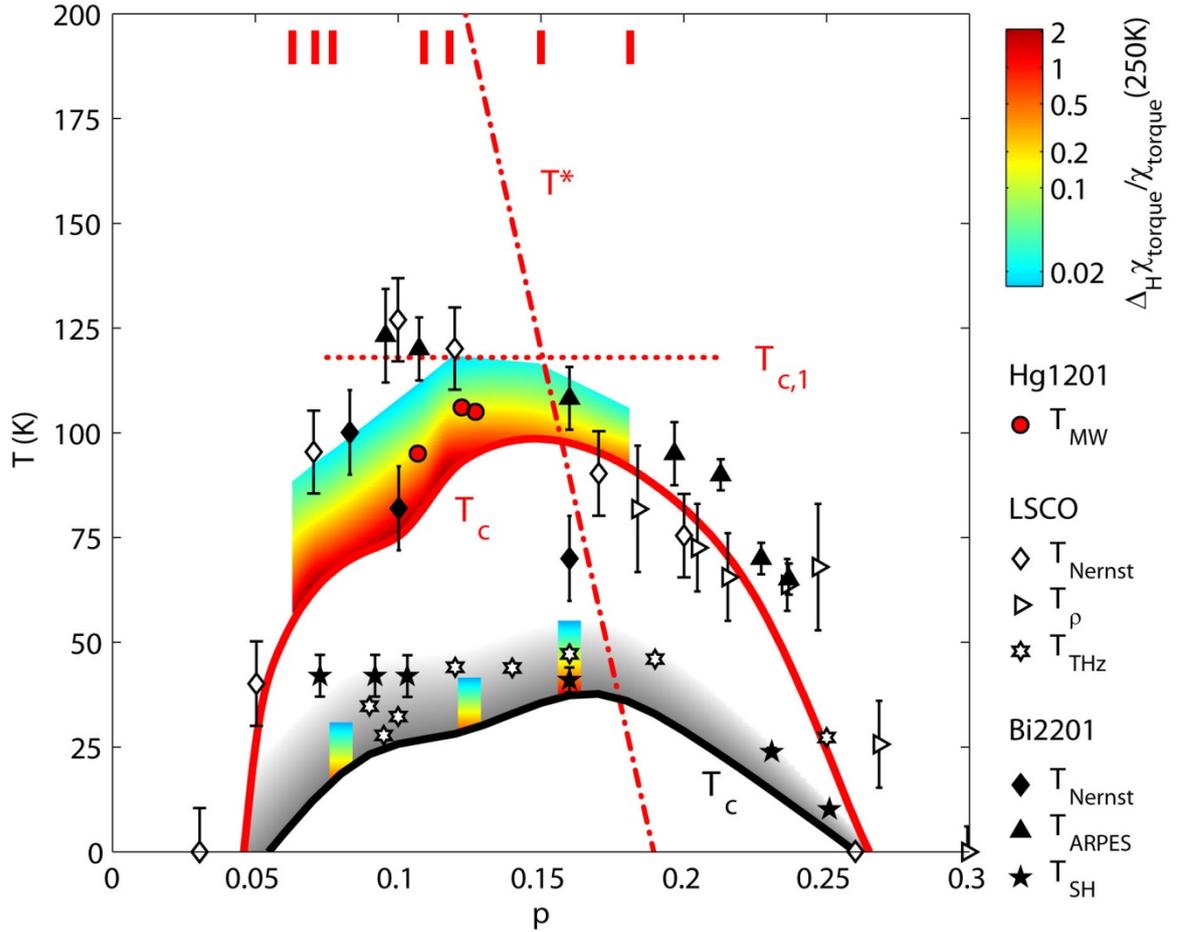

**Figure 4 | Phase diagram.** The color contour shows $\log_{10}(\Delta_H\chi_{torque}/\chi^{250}_{torque})$ for Hg1201 (same data as in Fig. 3d, but for wider temperature range), obtained from an interpolation of measurements of seven samples (indicated by vertical red bars), and for optimally-doped Bi2201 and underdoped LSCO ($p = x = 0.08$ and $0.125$). The grey shaded area indicates schematically the extent of SC fluctuations in LSCO and Bi2201. Similar to Hg1201 (torque and microwave[6] ($T_{MW}$)), the lower characteristic temperatures for LSCO (torque and teraherz[7] ($T_{THz}$)) and Bi2201 (torque and specific heat[9] ($T_{SH}$)) indicate that SC fluctuations are restricted to a narrow temperature range above $T_c$, and closely track $T_c$ with doping (shaded grey area). For LSCO, the high characteristic temperatures are from Nernst ($T_{Nernst}$)[1,2] measurements, whereas for Bi2201 they are from Nernst ($T_{Nernst}$)[2] and photoemission ($T_{ARPES}$)[5] measurements. For LSCO, the characteristic temperatures from dc magnetoresistance ($T_\rho$)[14] are also shown. See Fig. S4 for a direct comparison between current and prior[3,4] torque results for LSCO. The horizontal dotted line indicates $T_{c,1} = 118$ K, the maximum $T_c$ for single-layer ($n = 1$) cuprates, observed for optimally-doped Hg1201 at 20 GPa[27]. Overall, $T_{Nernst}$, $T_\rho$, and $T_{ARPES}$ for the low-$T_c^{max}$ compounds LSCO and Bi2201 agree rather well with $T_c(p)$ of Hg1201 on the overdoped side, and with $T_{torque}$ of Hg1201 on the underdoped side. These characteristic temperatures extend to $\sim T_{c,1}$ near optimal doping. See Fig. S2 for a similar analysis for double-layer cuprates. $T^*(p)$ deduced from neutron[17,18] and transport[6,16] measurements for Hg1201 extrapolates to zero at $p \approx 0.19$. The estimation of the hole concentrations for Hg1201 and Bi2201 is described in the SI. For LSCO, $p = x$ is the Sr concentration. The torque data for Hg1201 (for LSCO and Bi2201) were obtained with $H = 14$ T (with $H = 9$ T).

# Universal superconducting fluctuations and the implications for the phase diagram of the cuprates

G. Yu, D.-D. Xia, N. Barišić, R.-H. He, N. Kaneko, T. Sasagawa, Y. Li, X. Zhao, A. Shekhter, and M. Greven

## Supplementary Information

This document includes supplementary text, Figures S1-S5, and references.

**1) Methods**

The torque measurements were carried out with a high-sensitivity torque lever chip, using Quantum Design Inc. PPMS instruments, and detected through the resistance change of piezoresistive elements in the Wheatstone bridge on the chip. The sensitivity of the method is limited by the signal-to-noise ratio of the Wheatstone bridge. Whereas the signal is determined by the sample characteristics and sample mass and increases with field (for example, $\propto H^2$ in the paramagnetic state), the noise is determined by the chip characteristics and the measurement (sampling) time. From the signal-to-noise ratio of the field dependence of the effective moment $M_\text{torque}$ (Fig. 1b) we estimate that our method has a sensitivity comparable to the prior torque study of LSCO and Bi2201 that used a capacitive cantilever[3,4].

In Refs. 3-4, 'onset' temperatures of the fluctuation diamagnetism were estimated from torque data obtained in a 14 T field. We use both 9 T and 14 T fields to extract the SC fluctuations in our study. Sensitivity differences between the two methods are not expected to result in very different 'onset' fluctuation temperatures due to the exponentially decreasing diamagnetic signal (Figs. 2, S1-3). In light of our findings, we believe that the previous estimation of diamagnetism in a very broad temperature range is the result of an unwarranted subtraction scheme that relies on the assumption of a linear (in temperature) paramagnetic 'background' contribution.

Our analysis does not rely on any such assumption. Instead, the SC diamagnetism is directly identified through its (nonlinear) magnetic field dependence. We find that the superconducting diamagnetism at moderately high fields is characterized by an exponential decay of $\chi_\text{torque}$ (Figs. 2b & S1) with characteristic temperature $T_\text{fl}$. This unusual exponential behavior is revealed directly from $\chi_\text{torque}(T)$ in the vicinity of $T_\text{c}$, where the diamagnetic signal is significantly larger than the paramagnetic contribution. It is also revealed from the quantity $\Delta_\text{H}\chi_\text{torque}(T)$, which allows us to follow the exponential behavior over a wider temperature range, until it is limited by the sensitivity of our setup. Close to our sensitivity limit, the temperature dependence of the nonlinear magnetic signal $\Delta_\text{H}\chi_\text{torque}$ can be described by adding a small temperature-independent constant to the exponential decay. The need for this constant results from the small systematic errors in setting the angles, from the offset due to gravity, and from the magnetoresistance of the torque lever chip.

The high-temperature magnetization of the cuprates is linear in field due to Langevin diagmagnetism as well as Van Vleck and Pauli paramagnetism. In contrast, the superconducting diamagnetism above $T_\text{c}$ exhibits a complex field dependence. At very small fields, $H < H^*$, one expects a linear (in field) response, not observed in the field range of our experiment ($H > 0.1$ T). In our experiment, $|M(H)|$ features non-linear (in field) behavior at all temperatures where the SC diamagnetism is discernable. We emphasize that the maximum field in our measurements (9 T and 14

T) is sufficiently large to reveal the nonlinear diamagnetic contribution at the measureable 'onset' of the SC fluctuations. For example, Fig. 1d shows that $|\chi_{torque}(H_c)|$ monotonically decreases with increasing field. The nonlinear component of the diamagnetic response is still observable at the lowest field, $H_c \approx 2$ T, even at the temperature ($\approx 120$ K) where the nonlinear diamagnetic signal is just above the experimental sensitivity.

At temperatures close to $T_c$, we identify two types of magnetization behavior. As shown in Fig. 2a for Hg1201 UD67, we observe Ginsburg-Landau-like behavior[11] up to about $H_c = 1$ T: the diamagnetic magnetization is well described by a power-law $|M_{torque}| \propto H_c^{\beta}$ (i.e., $|\chi_{torque}| \propto H_c^{\beta-1}$) with $\beta \approx 0.6$. A similar behavior was observed earlier in the torque study of Bi2201 and Bi2212 (referred to as 'fragile London rigidity)[4], where $\beta$ was found to decrease from 1 to values less than 0.2 close to $T_c$. For Hg1201, just above $T_c$, we find $\beta$ close to 1/2, consistent with conventional Ginsburg-Landau fluctuations[11]. At higher fields, we find that the magnetization increases slower than $|H_c|^{0.5}$, and at even higher fields, it decreases with increasing field. All discussions of SC fluctuations in the main text refer to this latter regime, which we refer to as moderately-high-field regime.

We find that at such moderately high fields, the superconducting diamagnetism closely follows an exponential decay: $\chi_{torque} = \chi_0(H)\exp\{-(T-T_c)/T_{fl}(T,H)\} + \chi_p(T)$, where $\chi_p(T)$ is the field-independent paramagnetic contribution. Figure S1a&b demonstrates the exponential behavior in both $\chi_{torque}$ and $d\chi_{torque}/dT$ near $T_c$. At higher temperatures, as the SC diamagnetism weakens, this universal behavior is masked by the paramagnetic contribution $\chi_p(T)$. Nevertheless, the exponential behavior is evident over a wider temperature range (Fig. 2d and Fig. 3b&c) in $\Delta_H\chi_{torque}$, a quantity that removes the field-independent contribution $\chi_p(T)$. The temperature range of the exponential decay appears to be cut off only by the sensitivity of our method, i.e., a temperature independent constant in $\Delta_H\chi_{torque}$, as described above.

The characteristic decay rate $T_{fl}$ can also be extracted from a plot of $-1/d\ln(-\chi_{torque})/dT$ vs. $T$ (Fig. S1c). In Fig. 3d, we show $T_{fl}$ obtained with this method: we use the values of $-1/d\ln(-\chi_{torque})/dT$ at or very close to $T_c$, where the paramagnetic contribution $\chi_p(T)$ is negligible. At low magnetic field, where Ginsburg-Landau-like behavior is expected, $\chi_{torque}$ deviates from the exponential decay. At high field, the SC diamagnetism weakens, and the paramagnetic contribution $\chi_p$ becomes more significant. We find that $T_{fl}$ is determined most reliably in the field range $\mu_0 H_c = 2$ to 5 T, and we use $\mu_0 H = 3.1$-$3.4$ T in Fig. 3d. The temperature $T_{fl}$ has a weak tendency to increase with increasing field, which reflects the non-linear (in field) behavior of the SC diamagnetism. However, this does not affect the monotonic doping dependence of $T_{fl}$ for Hg1201 or the universal behavior for the three single-layer cuprates studied here.

We note that the exponential decay (with $T - T_c$) of the SC diamagnetism is consistent with an early high-frequency conductivity measurement of double-layer $Bi_2Sr_2CaCu_2O_{8+\delta}$ (Bi2212), which probed the short-time-scale phase fluctuation dynamics and found the phase correlation time to decrease exponentially above $T_c$, and to quickly reach the mean free time of carriers in the metallic state[S1].

For the contour plots in Figs. 3d and 4, we used the angular difference instead of the magnetic field difference of $\chi_{torque}$. In our experimental setup, the angular scans are less prone to systematic error. As all data were obtained with the same field, which minimizes the systematic error from the magnetoresistance of the chip.

Furthermore, data at different angles were obtained from a single scan, whereas changing the applied field requires a relatively long time and introduces more error due to a small signal variation with time. This signal variation observed over long periods of time is caused by the small drift in the resistivity of the elements of the bridge, and it is effectively cancelled in the angular difference, because data for $\theta = 45°$ and $67.5°$ were taken nearly simultaneously. We also performed full angular scans at intermediate temperatures (as shown in Fig. 1c) to evaluate the higher harmonics using Fourier analysis. Since a complete angular scan is relatively time consuming, it is not practical for a detailed measurement of the temperature dependence. Measurements at two angles were sufficient for the determination of the contour plots.

We normalize $\Delta_H \chi_{torque}$ by its value at a high temperature, because data were collected with multiple chips. The normalization removes the systematic error in the absolute value of $\chi_{torque}$ due to chip calibration as well as the error in the determination of the sample mass. We use the highest temperature (250 K) measured for sample OP96 for normalization. This temperature lies far above the onset of fluctuations. For all samples measured to 400 K, the monotonic temperature dependence of $\chi_{torque}$ was found to continue at higher temperatures (Fig. 1a), so that the contour plot in Figs. 3d and 4 do not depend on the particular choice of reference temperature.

In addition to the data presented in Fig. 3, in the contour plots (Figs. 3d and 4) we also show results for another three Hg1201 samples (two underdoped, with onset $T_c = 63$ and 89 K, and one overdoped, with onset $T_c = 90$ K). These newer results were obtained at a lower 9 T field using a 9 T (rather than 14 T) Quantum Design Inc. PPMS instrument. In order to combine the 9 T and 14 T data, we multiply the 9 T results by factors of 0.79 ($\Delta_H \chi_{torque}/\chi_{torque}^{250K}$) and 1.18 ($T - T_c$). This normalization leads to a nearly perfect match of the 14 T and 9 T data for the UD67 sample.

For the phase diagram in Fig. 4, the hole concentration of our Hg1201 samples are taken from the values of $T_c(p)$ obtained in ref. S2 (data available up to $p = 0.21$). The superconducting dome (thick red line) is extended on the overdoped side using the empirical relation $T_c/T_c^{max} = 1 - 82.6 (p - 0.16)^2$ (ref. S3). For Bi2201, $p$ is estimated from the above empirical relation. For LSCO, we use $p = x$.

**2) Torque measurements of Bi2201 and LSCO**
The optimally-doped Bi2201 crystal (onset $T_c \approx 35$ K) was grown by the traveling-solvent floating-zone technique[10]. The torque measurement (Fig. S2) gives a result closely similar to that for Hg1201. Unlike tetragonal Hg1201, Bi2201 has a long-range distorted crystal structure[S4]. However, since the planar lattice constants $a$ and $b$ differ by less than ~ 0.4%[S4], the difference between $\chi_a$ and $\chi_b$ is negligible compared to that between $\chi_a$ and $\chi_c$. Therefore, the same analysis can be applied to Bi2201.

As for Hg1201, the paramagnetic susceptibility of Bi2201 does not follow a linear temperature dependence, as evidenced by the fact that $d\chi_{torque}/dT$ is not constant throughout the entire measured temperature range. We observe a clear crossover at $T_{torque} \approx T_c + 20$ K $\approx 55$ K between two different temperature and field behaviors of the torque susceptibility (Fig. S2c&d). The near-$T_c$ regime exhibits a strong magnetic field dependence and a rapid exponential decay with $T-T_c$. Since this behavior is directly connected to the SC transition, it is unambiguously identified as SC diamagnetism. At higher temperatures, $\chi_{torque}$ exhibits a different temperature dependence and no measureable magnetic field dependence. In Fig. S2c, the deviation from a $T$-linear fit at higher temperature is plotted vs temperature. This plot clearly demonstrates the presence of a crossover, athough as emphasized in the main text, the apparent temperature and character of the crossover depend on the details of the subtraction. We conclude that the magnetic response defined as the deviation from an assumed high-temperature $T$-linear behavior in the prior torque studies[3,4] contains two components, the strongly field-dependent one within about 20 K above $T_c$ identified by our method, and a field-independent (within error) one. Consequently, the temperature scale extracted in the prior work corresponds to higher temperature physics. It is quite possible that, at intermediate and low hole concentrations, the strong temperature dependence of the magnetization in this high-temperature range is not an independent diamagnetic contribution, but rather results from a temperature-dependent suppression of the paramagnetic susceptibility associated with the opening of the pseudogap.

Our torque measurements for LSCO (Figs. 3c&d and S3) further confirm the conclusions reached for Hg1201 and Bi2201. Strong field dependence of $\chi_{torque}$ is only observable up to about 20 K above $T_c$. The characteristic temperatures obtained from the previous torque study are dramatically higher (Fig. S4) and associated with the unusual normal state.

In Fig. 4, we include our results for Bi2201 and LSCO in the comparison of characteristic temperatures obtained with different techniques for single-layer compounds. In evaluating Fig. 4, it should be kept in mind that, invariably, there exist sample specific differences (hole concentration estimates, chemical disorder) and different degrees of sensitivity to SC fluctuations among the experimental probes. For example, our Bi2201 sample has a $T_c$ that is 5 K higher than that of the nominally optimally-doped sample in the specific heat measurement[9].

**3) Additional considerations pertaining to Figure 4**
As shown in Fig. 4, our result for Hg1201 agrees well with microwave conductivity work[6]: the fluctuation onset identified with the latter technique corresponds to $\Delta\chi_{torque}/\chi_{torque}^{250K}$ = 0.1-0.2. Consistent with our torque measurements for the low-$T_c^{max}$ compound Bi2201, specific heat measurements indicate a low characteristic temperature for Bi2201[9] and a terahertz conductivity study of LSCO thin films shows that temporal phase correlations are observable only up to about $T_c$ + 20 K[7,8]. This consistency among results obtained with charge and magnetic probes reinforces the conclusion that measurable SC fluctuations in both low- and high-$T_c^{max}$ single-layer compounds do not extend very far above $T_c$. It is not clear at this time if a more quantitative comparison of the extent of SC fluctuations can be achieved for these different experimental probes. However, by using the same (highly sensitive) probe and applying the same standard, in the present work we are able to systematically compare low- and high-$T_c^{max}$ single-layer compounds. This reveals a universal fluctuation regime that is unconventional yet rather narrow, a result that is not apparent from a comparison of different experimental techniques or from the prior torque work.

A second class of experiments[1,2,5], including Nernst effect and photoemission, indicates that the low-$T_c^{max}$ compounds exhibit a second temperature scale that is dramatically higher than their $T_c$ (Fig. 4). Remarkably, these higher characteristic temperatures are close to $T_c(p)$ of the more ideal high-$T_c^{max}$ system Hg1201. Specifically, near optimal doping, $T_c$ of Hg1201 is comparable to the higher characteristic temperatures of LSCO and Bi2201, which in turn appear to be bound by the highest $T_c$ attainable in the single-layer ($n$ = 1) compounds, namely $T_{c,1}$ = 118 K observed in Hg1201 at high pressure[27]. On the overdoped side of the phase diagram, the higher characteristic temperatures of LSCO and Bi2201 approximately follow $T_c(p)$ of Hg1201. Consequently, Fig. 4 illustrates a seemingly universal doping-dependent temperature scale, approximately equal to $T_c(p)$ for Hg1201, which for the low-$T_c^{max}$ compounds is not associated with genuine SC fluctuations.

**4) Multilayer cuprates**
Our analysis of characteristic temperatures (Fig. S5) for the double-layer ($n = 2$) cuprates YBCO and Bi2212 supports these observations. Here, early terahertz conductivity[S1] and more recent Josephson tunneling[S5] experiments indicated that the SC fluctuation regime is unconventional, but rather narrow when compared to the interpretation of some of the results for LSCO and Bi2201 in Figs. 4 & S4.

Just as for the single-layer compounds (Fig. 4), the characteristic temperatures of the double-layer systems appear to be bound by the highest attainable $T_c$ ($T_{c,2} = 154$ K at high pressure) observed in Hg1212 at high pressure[27] (Fig. S5).

Introducing the effective transition temperature $t_c^{eff} \equiv T_c^{max}/T_{c,n}$, we find for the low-$T_c^{max}$ compounds Bi2201 and LSCO that $t_c^{eff} = 0.30\text{-}0.32$, whereas $t_c^{eff} = 0.82$ for Hg1201 and Hg1212 at ambient pressure. Double-layer YBCO and Bi2212 have intermediate values of $t_c^{eff} = 0.60\text{-}0.62$.

Since both YBCO and Bi2212 exhibit intermediate $t_{c,eff}$ values, the distinction between low and high characteristic temperatures is less clear than for single-layer LSCO and Bi2201 (Fig. 4). Nevertheless it can be seen that, as for the single-layer compounds, the microwave result (YBCO)[S6] is close to $T_c$, whereas the characteristic temperatures from magnetoresistance (YBCO)[24], photoemission (Bi2212)[5] and Nernst effect (Bi2212)[2] are high.

One exception is the Nernst result for YBCO, which lies close to $T_c$. It therefore appears that the Nernst effect is neither a straightforward probe of SC fluctuations (vortices) nor of local pair formation or charge-density-wave order, since these phenomena may have comparable contributions[25]. Similarly, whereas the mean-field transition temperature $T_{MF}$ derived from specific heat measurements of Bi2212[S7] agrees well with the high end of the other characteristic temperatures (Fig. S5), for Ca-doped YBCO[S7] $T_{MF}$ is intermediate (not shown), and for single-layer Bi2201[9] it corresponds to the lower characteristic temperature ($T_{SH}$ in Fig. 4).

The pairing temperature $T_\Delta$ from STM for optimally and overdoped Bi2212 (defined as the temperature at which 10% of the cleaved sample surface is gapped[S8]) is in remarkably good agreement with the high characteristic temperature from bulk probes (Fig. S5). This would seem to be the result of the fact that STM is capable of detecting minority regions of very short length scale, which are undetectable in most bulk measurements. Of course, for a more quantitative comparison, one would also have to consider differences in disorder in the bulk and on the cleaved surface.

Finally, we note that for triple-layer ($n = 3$) $Bi_2Sr_2Ca_2Cu_3O_{10+\delta}$, the Nernst effect for a $T_c = 109$ K sample gave an 'onset' temperature of about 135 K[2], the same value as the $T_c^{max}$ of $HgBa_2Ca_2Cu_2O_{8+\delta}$ (Hg1223) at ambient pressure, but about 20% below the high-pressure value of $T_{c,3} = 164$ K for Hg1223, the record superconducting transition temperature of the cuprates[27]. For Hg1223, $t_c^{eff} = 0.82$, as for Hg1201 and Hg1212.

**5) Disorder in the cuprates**

The cuprates are intrinsically disordered and exhibit considerable variation in local hole concentration, Cu-O bond angles and bond distances[10,S9-13]. Even stoichiometric YBa$_2$Cu$_3$O$_7$ (YBCO) is not as 'clean' a system as often assumed: the local hole concentration was shown by NMR to vary substantially (~ 20%), and EXAFS data indicate a relatively large local strain of the CuO$_2$ planes[S12]. The crystal structures of the cuprates consist of an intergrowth of Cu-O layers of fixed oxygen composition and oxide layers with (typically) variable oxygen concentration. In general, there exists a bond-length mismatch across the interlayer interface, which corresponds to a deviation of the geometric tolerance factor from unity and typically leads to lower than tetragonal structural symmetry[S14-15]. The bond-length mismatch is further accommodated by local deviations from the average crystal structure. Additional sources of disorder include substitutional doping (as in LSCO), off-stoichiometry in the intervening layers (in most compounds), and structural domain boundaries in compounds with lower than tetragonal symmetry.

In comparison to Hg1201, the low-$t_c^{eff}$ single-layer compounds LSCO and Bi2201 feature disorder sites closer to the Cu-O planes[10], a larger local elastic strain[S12], and a lower structural symmetry, and hence nanostructured disorder in the Cu-O planes is expected to be more prevalent. Indeed, LSCO has been shown to exhibit a particularly large local hole distribution[S13].

A SQUID microscopy study of LSCO found that diamagnetic islands first form locally at high temperature, and then grow in area as $T_c$ is approached from above[S16]. While this observation would seem to suggest a substantial distribution of local $T_c$ values, these findings were not reproduced in studies of surfaces of bulk samples[S17]. Moreover, even in the case of low-$t_c^{eff}$ compounds, the SC transition widths in high-quality samples are found to be very narrow from bulk measurements, which rules out a substantial spatial variation of $T_c$[S18]. Consequently, while there may be considerable differences in the disorder morphology of the different cuprates[S19], with nanostructured disorder more prevalent in low-$t_c^{eff}$ compounds or on Bi2212 surfaces studied by STM and photoemission, disorder in the cuprates is inherently homogeneous and local rather than granular.

In the low-$t_c^{eff}$ cuprates, the situation appears to be similar to that of the conventional superconductor MgB$_2$ upon introducing point defects through neutron-irradiation[S20], which retains a sharp SC transition without a substantial spatial distribution of $T_c$. We note that for YBCO it has been found that electron irradiation[24,S21] and substitution of Y with Pr[S22] leads to a suppression of $T_c$ that is much stronger than the relatively weak reduction of the characteristic temperatures from Nernst effect[S21,S22] and magneto-transport measurements[24].

Non-local (singular) interactions can give rise to an unconventional softening of phase stiffness in a temperature range below the underlying SC pairing temperature. In compounds like LSCO and Bi2201, there might in principle exist rare regions with a local electronic environment that resembles the more ideal situation for $t_{c,eff} = 1$ and in which the pairing energy is high, giving rise to isolated pairs that are completely incoherent. In an STM study of double-layer Bi2212, a (static) spatial distribution of local gaps was revealed well above $T_c$[S8]. In subsequent theoretical work, it was proposed that the observed electronic inhomogeneity is due to a density wave of $d$-wave Cooper pairs without global phase coherence[S23]. As demonstrated in Fig. S5, at and above optimal doping the temperature below which the local gaps are found to appear is in rather good agreement with the higher characteristic temperatures for the $n = 2$ compounds. An alternative interpretation of the observed universal high temperature scale is that strong disorder together with singular interactions leads to charge-ordering tendencies in the low-$t_c^{eff}$ and medium-$t_c^{eff}$ compounds[25], and to complex non-universal behavior in a broad temperature range above $T_c$[S24]. Indeed, recent X-ray experiments[28,29] uncovered charge-density wave correlations in underdoped YBCO. These charge correlations were observed below about 150 K, i.e., below $t_c^{eff} \approx 1$ (see Fig. S5), and they were found to compete with the SC state.

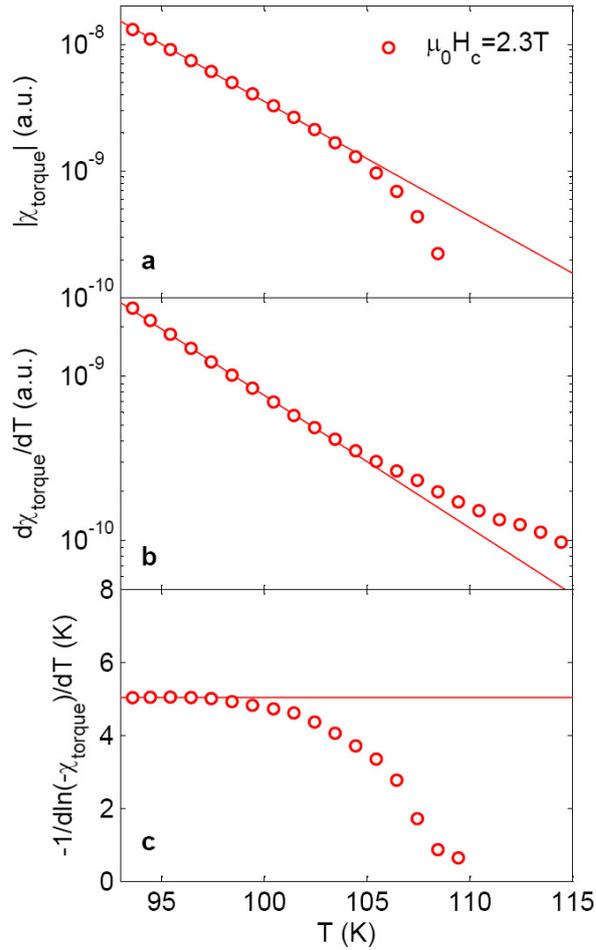

**Figure S1 | Extraction of characteristic superconducting fluctuation temperature $T_{fl}$ from $\chi_{torque}$ for Hg1201 ($T_c$ = 89 K). a,** Near $T_c$, for magnetic fields outside the Ginzburg-Landau-like regime ($H_c > 1$ T ; see Fig. 2d), the magnitude of $\chi_{torque}$ is well described by a simple exponential decay, $\chi_{torque} \propto \exp\{-(T-T_c)/T_{fl}\}$ (solid line), allowing the definition of the characteristic fluctuation temperature $T_{fl}$. The deviation from the exponential behavior above ~105 K is due to the increasing relative importance of the paramagnetic normal-state contribution. **b,** The temperature derivative $d\chi_{torque}/dT$ demonstrates the same exponential behavior as in **a** and is less sensitive to the paramagnetic contribution. **c,** Equivalently, the logarithmic derivative $-1/d\ln(-\chi_{torque})/dT$, which corresponds to the ratio of the quantities in panels **a** and **b**, equals $T_{fl}$ near $T_c$. The deviation from the constant value ($T_{fl}$) is due to the paramagnetic contribution that dominates at higher temperature. As shown in Fig. 3d, the doping (and compound) dependence of $T_{fl}$ is small. We note that $T_{fl}$ weakly increases with increasing $H_c$, but that this does not affect the fact that $T_{fl}$ gradually decreases with increasing doping. While $T_{fl}$ is best defined from $\chi_{torque}$, the difference $\Delta_H\chi_{torque}$ used in Figs. 2-4 and Figs. S2-S4 demonstrates the exponential decay to higher temperature and gives a better estimate of the measurable extent of superconducting fluctuations.

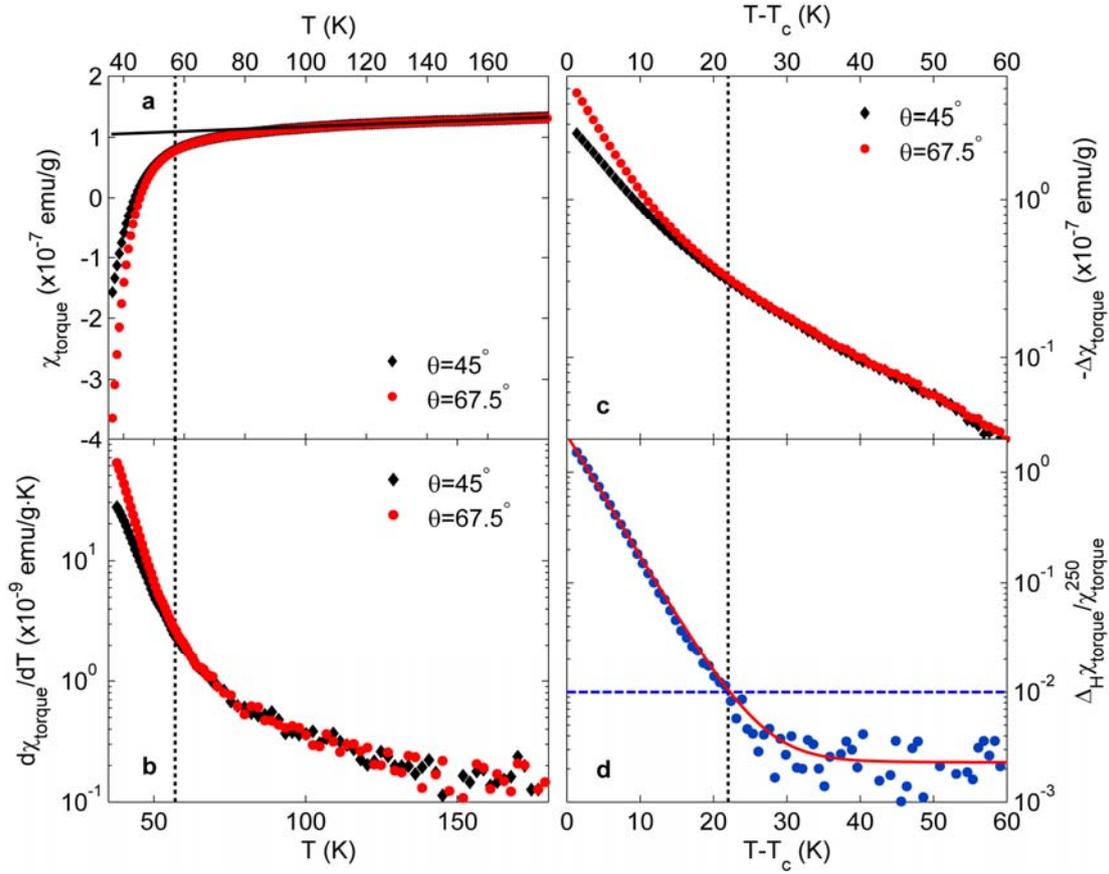

**Figure S2 | Torque susceptibility for optimally-doped Bi2201 ($T_c \approx 35$ K).** Data were taken with a 9 T field. **a,** $\chi_{torque}$ obtained at $\theta = 45°$ and $67.5°$, i.e., at different values of $H_c$, the magnetic field component perpendicular to the CuO$_2$ planes. The black solid line is a linear fit to the data between 100 K and 180 K. The deviation from an apparent high-temperature $T$-linear behavior was previously interpreted as due to fluctuation diagmagnetism[3,4]. However, a field-dependent diamagnetic response is noticeable only at the lower temperature indicated by the vertical dashed line. **b,** The temperature derivative d$\chi_{torque}$/d$T$ changes continuously at all temperatures. **c,** $\chi_{torque}$ with the $T$-linear fit (defined in **a**) subtracted. The data clearly show two distinct temperature regimes. The near-$T_c$ regime is characterized by a strong magnetic field dependence and a rapid exponential decay of the diamagnetic magnetic response with increasing temperature. The higher temperature regime exhibits no field dependence within our sensitivity limit. **d,** $\Delta_H \chi_{torque}/\chi^{250}_{torque}$ from the difference between data taken at $\theta = 45°$ and $67.5°$. The red solid line is a fit to the form $\exp\{-(T-T_c)/T_\Delta\}$ + *constant*. The blue horizontal dashed line indicates the limit $\Delta_H \chi_{torque}/\chi^{250}_{torque} = 0.01$ below which the diamagnetic signal due to SC fluctuations can no longer be reliably discerned. The vertical dashed line indicates the corresponding characteristic temperature.

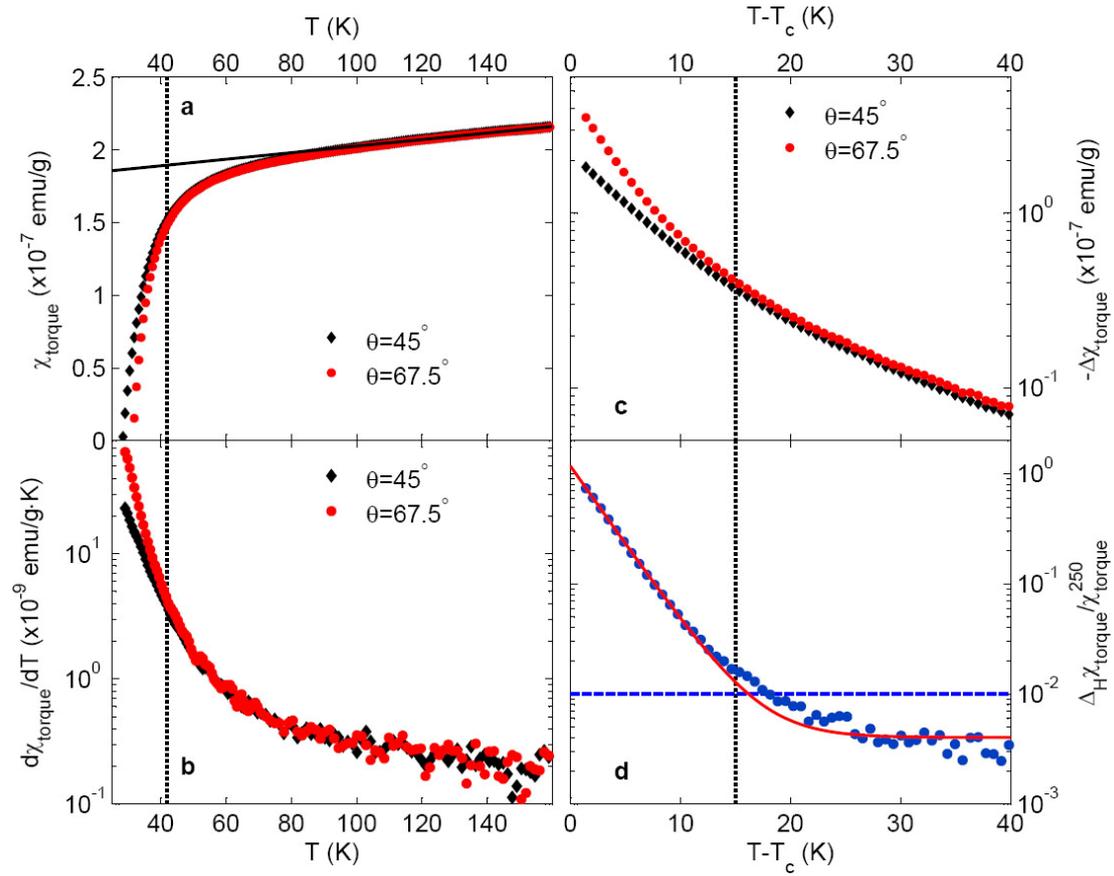

**Figure S3 | Torque susceptibility for La$_{1.875}$Sr$_{0.125}$CuO$_4$ ($T_c \approx$ 27 K).** Data were taken with a 9 T field. **a,** $\chi_{\text{torque}}$ obtained at θ = 45° and 67.5°, i.e., at different values of $H_c$, the magnetic field component perpendicular to the CuO$_2$ planes. The black solid line is a linear fit to the data between 100 K and 180 K. The deviation from an apparent high-temperature $T$-linear behavior was previously interpreted as due to fluctuation diagmagnetism[3,4]. However, a field-dependence of the magnetic response is noticeable only at much lower temperature. **b,** Temperature derivative d$\chi_{\text{torque}}$/d$T$ changes continuously (all the way to 350 K, the highest temperature of the measurement), indicating that there is no true $T$-linear range at this doping level. **c,** $\chi_{\text{torque}}$ with the $T$-linear fit (as defined in **a**) subtracted. The data show two distinct temperature regimes. The near-$T_c$ regime is characterized by a strong magnetic field dependence and a rapid exponential decay of the magnetic response with increasing temperature. The higher-temperature regime exhibits no field dependence within our sensitivity limit. **d,** $\Delta_H\chi_{\text{torque}}/\chi^{250}_{\text{torque}}$ from the difference between data taken at θ = 45° and 67.5°. The red solid line is a fit to the form exp{−($T$−$T_c$)/$T_\Delta$} + *constant*. The blue horizontal dashed line indicates the limit $\Delta_H\chi_{\text{torque}}/\chi^{250}_{\text{torque}}$ = 0.01 below which the diamagnetic signal due to SC fluctuations can no longer be reliably discerned. The vertical dashed line in all panels indicates the corresponding characteristic temperature.

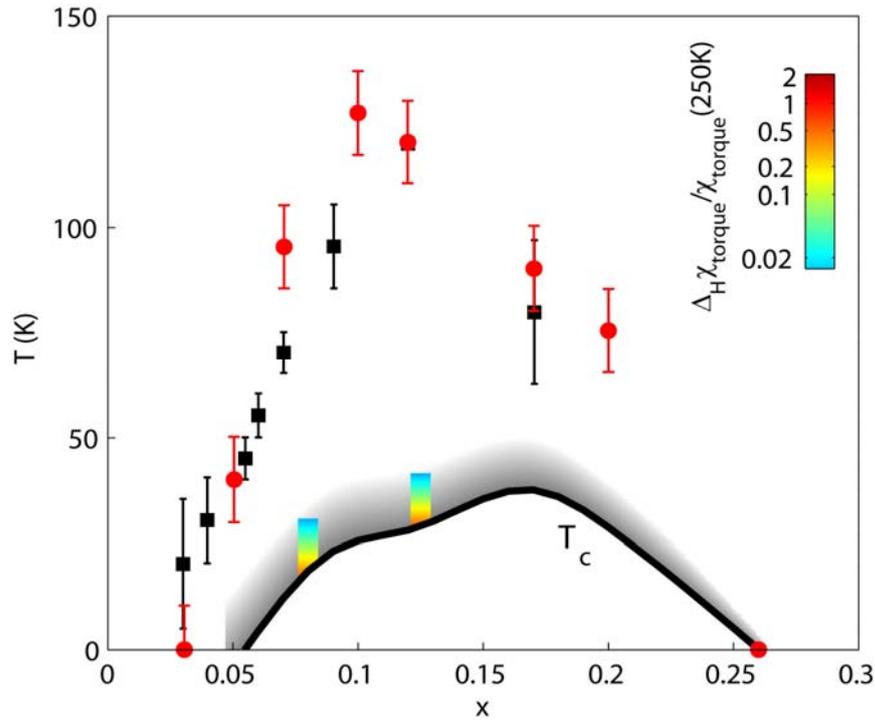

**Figure S4 | Comparison with prior torque and Nernst work for LSCO.** Pioneering Nernst effect measurements[1,2] (circles) yielded very high characteristic temperatures and were interpreted as indicative of SC fluctuations. Subsequent torque data[4] (squares) were argued to be consistent with the Nernst results. In contrast, our torque data imply a dramatically smaller SC fluctuation regime. The color contour shows $\log_{10}(\Delta_H \chi_{torque}/\chi^{250}_{torque})$ for the $x = 0.08$ and $0.125$ samples (see also Fig. 4). The exponentially decreasing diamagnetic signal is below our detection limit for temperatures above about $T_c + 20$ K.

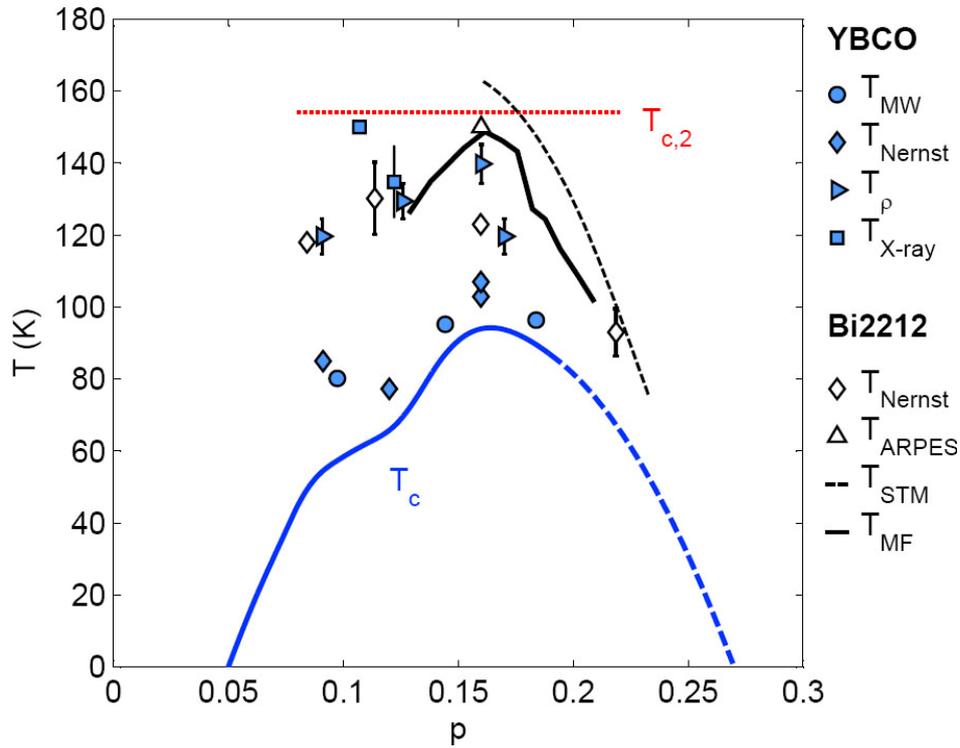

**Figure S5 | Characteristic temperatures of the double-layer cuprates.** The characteristic temperatures for YBCO (from Nernst effect $(T_{Nernst})$[2,S21,S25], dc transverse magnetoresistance $(T_\rho)$[24], and the observation of charge-density-wave order from X-ray scattering $(T_{X-ray})$[28,29]) and for Bi2212 (from Nernst effect $(T_{Nernst})$[2] and photoemission spectroscopy $(T_{ARPES})$[5] and STM $(T_{STM})$[S8]). All these characteristic temperatures are bound by $T_{c,2}$, the highest attainable $T_c$ in the double-layer cuprates (attained for Hg1212 at high pressure). For Bi2212, the mean-field transition temperature $T_{MF}$ derived from specific heat measurements[S7] is also shown. Even more so than photoemission, which probes a few unit cells near the cleaved sample surface, STM is a surface probe. Although bulk and surface disorder effects are expected to be different, above optimal doping the characteristic temperature associated with the inhomogeneous opening of local gaps is in rather good agreement with the other results in the figure. The blue solid line indicates $T_c(p)$ for YBCO[S26], which is extended to higher doping using the empirical formula $T_c/T_c^{max} = 1 - 82.6\,(p - 0.16)^2$ (ref. S3).